\documentstyle[12pt,epsf]{article}
\newcommand{\beq}{\begin{equation}}
\newcommand{\dd}{\partial}
\newcommand{\eeq}{\end{equation}}
\newcommand{\bea}{\begin{eqnarray}}
\newcommand{\eea}{\end{eqnarray}}

\newcommand{\phib}{\bar{\phi}}

\newcommand{\Ql}{\tilde{Q}_{_L}}
\newcommand{\tr}{\tilde{t}_{_R}}

\begin{document}

\baselineskip 7.5 mm

\def\thefootnote{\fnsymbol{footnote}}

\begin{flushright}
\begin{tabular}{l}
UPR-709-T \\
August, 1996 
\end{tabular}
\end{flushright}

\vspace{12mm}

\begin{center}

{\Huge \bf   Is the vacuum stable?
}
\\

\vspace{8mm}

\setcounter{footnote}{0}

Alexander Kusenko\footnote{ email address:
sasha@langacker.hep.upenn.edu; address after October 1, 1996:
Theory Division, CERN, CH-1211 Geneva 23, Switzerland
} and 
Paul Langacker\footnote{ email address: pgl@langacker.hep.upenn.edu}
\\ 
Department of Physics and Astronomy, University of Pennsylvania \\ 
Philadelphia, PA 19104-6396 \\

\vspace{20mm}

{\bf Abstract}
\end{center}
The experimental data, as well as theoretical considerations
allow (and, in some cases, require) the Universe at present to rest in
a false vacuum, whose approximate stability imposes constraints on the
model parameters.  Under very general and mild conditions, the Universe 
would have ended up in the standard vacuum even if the potential has deeper 
minima, provided there was a period in which the temperature was 
$\stackrel{>}{_{\scriptstyle \sim}} 1$~TeV.  In many cases,
the zero temperature tunneling 
rate is much smaller than the inverse age of the Universe. 
Future experiments may reveal that the physical vacuum is not
entirely stable.  Implications for the cosmological constant are discussed.

\vfill

\pagestyle{empty}

\pagebreak

\pagestyle{plain}
\pagenumbering{arabic}
\renewcommand{\thefootnote}{\arabic{footnote}}
\setcounter{footnote}{0}

\pagestyle{plain}

The question of the stability of the physical vacuum is  of fundamental
importance and has obvious scientific and cultural implications. 
If the Universe at present is in a metastable state with a lifetime of
order billions of years, then its future evolution may be determined by a
catastrophic phase transition into a lower-energy state.  The past history 
of the big bang Universe would also be elucidated if we knew that it ended
up in a false vacuum.  

This fundamental question has received a lot of attention since
the advent of supersymmetry, which allows for a complicated structure of
false vacua at the electroweak scale.  When  future experiments 
unveil the structure of the potential at the scale $ \sim 1$ TeV, we may
learn about the existence of states with  lower energy density than
that of the present vacuum.  It is timely, therefore, to examine this 
issue in the light of recent theoretical developments.

A generic feature of theories with (softly broken) supersymmetry is a
scalar  potential,  $V(\phi)$, that depends on a large number of scalar
fields $\phi=(\phi_1,...,\phi_n)$.   For this reason, the scalar potential
of the  MSSM, unlike that of the Standard Model, may have a number of local
minima characterized by different gauge symmetries.   In particular, the
supersymmetric partners of quarks, $\tilde{Q_{_L}}$ and $\tilde{q_{_R}}$, 
may have non-zero vev in some minima, where the tri-linear terms 
$A H_2 \tilde{Q_{_L}} \tilde{q_{_R}} $  and $\mu H_1 \Ql \tr$  are large and
negative (here $H_{1,2}$ denote Higgs fields, $A$ is the SUSY breaking
parameter, and $\mu$ is the coefficient of $H_1 H_2$ in the
superpotential).  These color and charge breaking (CCB) minima may be
local or global, depending on the values of the MSSM parameters.  There
might also be directions along which the effective potential is unbounded
from below\footnote{ In such a case the effective potential would
presumably receive large positive contributions near the Planck scale or
next physics threshold, so that the apparently UFB direction would really
correspond to a very deep minimum.} (UFB), in which case all the minima are
local. 

Any of these local minima may serve as the ground state for the Universe at
present, provided that the lifetime of the metastable state
is large in comparison to the age of the Universe.  The latter is plausible
\cite{tw,chh,kls} because the tunneling rate in quantum field theory is
naturally suppressed by the exponential of a typically large dimensionless
number, the saddle point value of the Euclidean action \cite{tunn}.  
The probability of the first-order phase transition does not necessarily
increase with the depth of the true vacuum relative to that of the false
vacuum.  In fact, there is a natural scale, the so called ``escape point'',
which serves as an ultraviolet cutoff, such that physics at larger energy
scales does not affect the tunneling probability (in semiclassical
approximation).  This non-perturbative decoupling allows one to treat the
UFB directions on the same footing as the very deep minima of the
potential~\cite{kls}. 

It is possible to detect the metastability of the false
vacuum empirically, at least in principle.  The most direct way is 
to infer the structure of the scalar potential near its minimum 
from future collider experiments.  Renormalizability forces $V(\phi)$
to be a fourth degree polynomial, up to the terms that are suppressed by the
powers of some large scale.  Supersymmetry imposes further constraints on 
$V$.  The global structure of the potential can then be derived 
in principle from its local properties near the minimum, possibly revealing
new deeper minima.  

One can imagine a Gedanken experiment to determine directly
whether the vacuum is true or false.  The metastability implies that the
scalar potential has a non-zero imaginary part \cite{tunn}.  Therefore, 
in principle, one could expect small breaking of T and CP, as well as
SUSY, were they not already broken.  However, if a metastable vacuum has
existed for $\tau_{_U}=$ 10 billion years, then any effects of the
metastability \cite{ak3}  would be characterized by the scale $<
1/\tau_{_U} \sim 10^{-33}$ eV, beyond any hope of being observable.
These effects may, however, play a role in the early Universe in the false
vacuum at the brink of a first order phase transition~\cite{ak3}.

Different minima of the scalar potential are characterized by different
values of the cosmological constant.  However, contrary to recent 
claims \cite{cd,bbc}, the cosmological constant problem is just as severe
in the stable vacuum as it is in a metastable one.  
In fact, in a large class of 
Unified theories with unbroken local supersymmetry 
the cosmological constant can be fine-tuned to zero in any of the {\it
local} minima, but not in the global minimum \cite{w}, where the vacuum
energy does not vanish even at the expense of naturalness.  
The cosmologically acceptable false vacuum is absolutely stable in this case
because of the Coleman --- De Luccia \cite{cdl} suppression of tunneling.  

Similarly, many models with dynamical breaking of supersymmetry  predict the
existence of a global supersymmetry preserving 
vacuum in addition to the standard vacuum with the correct pattern of gauge
and supersymmetry breaking (see Ref. \cite{r} and references therein).
Since this may be the way SUSY is broken in the real world, the idea that
we live in a metastable vacuum appears rather plausible. 

Just about any of the proposed solutions \cite{w_r} to the cosmological
constant problem (none of which is fully satisfactory) would work equally
well in a local, or global minimum.   Clearly, all the anthropic
considerations depend exclusively on the physical properties of the vacuum
and its accessibility (discussed below) in the course of the evolution of
the Universe.  An intriguing possibility is some kind of an adjustment
mechanism, similar to those discussed in Refs. \cite{w_r,a}, which would
naturally set the cosmological constant to zero regardless of its initial
value.  A mechanism of this sort would also be largely  independent of
whether the vacuum is true, or false.  Finally, the cosmological constant
may vanish because of some exact symmetry \cite{w_r,wtn}, which  
may or may not require the system to be in its true ground state.  

Therefore, based on all the empirical and theoretical considerations, if
the scalar potential has more than one minimum, the physical vacuum  
at present should be determined by the evolution of the Universe. 

We concentrate on the TeV-scale CCB minima.  Nontrivial minima of the
scalar potential usually disappear at temperatures $T$ much larger than the
mass scales in the potential.  In a general quantum field theory with
several scalar fields, it is sometimes possible to find small regions of
parameter space which evade this result, {\it e.\,g.}, for a gauge symmetry
to be broken by the finite-temperature corrections \cite{lpi}.  However, 
this does not occur in softly broken SUSY theories, because  the mass
matrix of scalars always receives a positive definite 
contribution \cite{m} $\Delta M^2(T)$ proportional to $T^2$, 
as long as $T$ is larger than all the masses in the theory.  (The latter 
also ensures that the high-temperature expansion, used here implicitly, 
is well defined.)  At the temperature $ \sim 1$ TeV, the supertrace 
sum rule implies that the scalar mass matrix, in the basis of eigenvectors 
of $\Delta M^2(T)$, receives the following contribution \cite{ce}: 
 
\beq
M^2_{ij} \equiv \frac{\dd^2 V(\phi,T)}{\dd \phi_i \dd \phi_j}
\rightarrow M^2_{ij}+ \Delta M^2(T) = 
M^2_{ij}+ \frac{\delta_{ij}}{8} 
\left \{ \sum_{kl} |W_{ikl}|^2+4 \sum_a g^2_a \; C_a(R_i) \right \} T^2,
\label{thermal}
\eeq 
where $W_{ikl}=\dd^3 W/\dd \phi_i \dd \phi_j \dd \phi_k$
are the Yukawa couplings in the superpotential $W$, $g_a$ are the gauge
couplings and $C_a(R_i)=(T^a)^2_{ii}$ are the corresponding Casimir
invariants. 

Since the one-loop thermal corrections in equation (\ref{thermal}) are
positive, the symmetry will be restored at a sufficiently high value of
the temperature; in the case of the electroweak scale CCB minima, this 
will happen at $T > M_{_{S}} \sim $ few TeV.  If the reheating temperature  
after inflation is higher than  
a few TeV, then the $SU(3)\times SU(2)\times U(1)$-symmetric phase will be
the starting point of the electroweak phase transition.  The same is also 
true \cite{rrv} if the preheating via parametric resonance decay of the
inflaton \cite{linde} follows inflation.   

As was argued in Ref. \cite{kls}, at the temperature just above the
electroweak phase transition there are two possibilities: (i) that of a
likely nearly-second-order phase transition into a standard minimum, or  
(ii) a far less probable first order phase transition
into a CCB minimum.  The difference is that transition (i) is associated
with $T$-dependent scalar masses, and is second order at tree level, while
the CCB minima are driven by large cubic terms in the effective potential
and are usually strongly first order\footnote{
The requirement that the second-order transition into
a CCB minimum be impossible imposes relatively weak constraints on the
scalar mass terms \cite{kls}. }. Therefore, the evolution of the early 
Universe, under some very general and not very restrictive conditions,
appears to favor the color and charge conserving minimum.  This 
{\it a~priori} unexpected benefit of the low-energy supersymmetry is
remarkable and intriguing.  Similar arguments apply to the UFB directions. 

For the Universe to remain in a metastable state, the latter
should have a lifetime of order billions of years.  The tunneling
rate can be evaluated in the semiclassical approximation
\cite{tunn,prefactor} and is proportional to $\exp (-S[\phib])$, where 
$ S[\phib] $ is the Euclidean action of the so called ``bounce'',
$\phib(x)$, a solution  of the classical Euclidean field equations.  
In practice, however, finding $\phib(x)$ numerically is very difficult (or
nearly impossible), especially in the
case of a potential that depends on more than one scalar field. This is
because $\phib(x)$ is an unstable solution, as it must be to be a saddle
point of the functional $S[\phi]$.  An effective alternative to solving the
equations of motion is to use the method of Ref. \cite{ak1}.  The idea is
to replace the action $S$ with a different functional, $\tilde{S}$, for
which the same solution, $\phib(x)$, is a minimum, rather than a saddle 
point. Then $\phib(x)$ can be found numerically using a
straightforward relaxation technique to minimize $\tilde{S}$.

A number of constraints arise from requiring the standard vacuum to be the
global minimum of the potential \cite{cd,bbc,ccb}.  However, it is clear from
the discussion above that such a requirement is too extreme and
unjustified.  It was shown in Ref. \cite{kls} that allowing metastability 
relaxes the constraints on the tri-linear couplings for the third
generation of squarks, those associated with large Yukawa couplings.

At the same time, cosmological considerations leave no room for 
constraints on the MSSM parameters from 
the CCB minima related to smaller Yukawa couplings.   
As an example, let us consider the constraints on flavor-violating
terms advocated in Ref. \cite{cd}.  It was argued that the tri-linear
coupling $A$ that enters in the soft SUSY-breaking term $A H e \tau $ cannot
exceed the bound  

\beq
|A|^2 \leq y_\tau^2 (m^2_e+m^2_\tau+m^2_{_H}),
\label{bound}
\eeq
which is more stringent than the experimental limit from the absence of
FCNC. Here $y_\tau$ is the $\tau$ Yukawa coupling; $e,\ \tau$, and $H$ are the 
left-handed electron, right-handed $\tau$, and Higgs fields, respectively;
and $m$'s are their mass terms.  The constraint (\ref{bound}) results from 
the requirement that there be no minimum of $V$ in the direction
$|e|=|\tau|=|H|=a$, along which 

\beq
V=(m^2_e+m^2_\tau+m^2_{_H}) a^2 - 2 A a^3 + y_\tau^2 a^4 
\label{V}
\eeq

If the potential (\ref{V}) has a global minimum at $a>0$, this minimum will
disappear at sufficiently high temperature.  By virtue of the same argument
as in Ref. \cite{kls}, we expect a nearly-second-order transition into the
standard electroweak vacuum to occur before the tunneling into a
charge-breaking minimum can take place.  In fact, since the Yukawa coupling
is small, the tunneling probability at both finite and zero temperature 
will be suppressed by the exponential of $(-1/y_\tau^2)$.  This makes 
tunneling practically impossible at $T=0$, as well as at $T>0$. 

The CCB minima associated with small Yukawa couplings appear to be
completely isolated and unreachable except if the scalar mass matrix 
in the $SU(3)\times SU(2)\times U(1)$-symmetric phase acquires a negative
eigenvalue at $T>T_c$, where $T_c$ is the electroweak phase transition
temperature.   The latter requires  mass-squared terms which are large in
magnitude and opposite in sign \cite{kls}, not a generic set of
parameters by any means.  In addition, the F-term contribution of the form 
$|y H \tilde{f}|^2$ to the sfermion $\tilde{f}$ mass matrix at the
standard vacuum must be larger \cite{kls} than $c \: T^2_c$, where $c \sim
1$.  Since $|y H|^2 = m_f^2$, the corresponding fermion mass, the condition 
$|y H|^2 =m_f^2 \ > \ c \: T_c^2$ cannot be satisfied for light fermions.  
Therefore,  the CCB minimum associated with a small Yukawa coupling would
not be populated either before (via a second-order 
transition or thermal tunneling)  or after (by means of tunneling) the
breaking of the electroweak symmetry.  We conclude that cosmological
considerations do not support a bound of the type (\ref{bound}).

Tunneling into a UFB direction can be treated on the same footing with that
into a deep CCB minimum \cite{kls}.  Therefore, all the above arguments
apply to the UFB bounds provided that the UFB directions are lifted in the
early Universe and the vev's are driven towards the $SU(3)\times
SU(2)\times U(1)$-symmetric minimum.  A combined effect of the effective
mass terms proportional to the Hubble parameter, finite-temperature
contributions and supersymmetry breaking by the large density scalar
condensates during the preheating is to restore the electroweak symmetry 
at temperatures $\sim 1$ TeV, unless the reheating temperature is
unacceptably low \cite{rrv}.  Thus no constraints result from considering
the UFB directions related to the small Yukawa couplings.

In summary, we find that only the CCB minima associated with the large 
Yukawa couplings provide some constraints on the MSSM parameters, 
as one requires the standard model-like vacuum to be stable with respect to 
tunneling.   

On the other hand, the existence of vacua with negative energy
density is acceptable, natural and, in some cases, desirable from a
theoretical  point of view, and is in agreement with all the present data.  
It is possible, therefore, that, as the future experiments yield
information about the scalar potential at the TeV energy scale, we may
learn that the Universe is resting in a false vacuum, --- a discovery of
Kopernikan importance with far-reaching scientific and cultural 
ramifications. 

We thank the Aspen Center for Physics for its hospitality.
This work was supported by the U. S. Department of Energy Contract No.
DE-AC02-76-ERO-3071.


\begin{thebibliography}{99}

\bibitem{tw} M. S. Turner and F. Wilczek, Nature {\bf 298} (1982) 633.

\bibitem{chh} M. Claudson, L. J. Hall and I. Hinchliffe, Nucl. Phys. 
{\bf B228} (1983) 501.

\bibitem{kls} A.~Kusenko, P.~Langacker and G.~Segr\`{e},  Phys. Rev. 
{\bf D}, in print, UPR-0677-T (hep-ph/9602414).

\bibitem{tunn} %
M. B. Voloshin, I. Yu. Kobzarev and L. B. Okun', Yad. Fiz.
{\bf 20} (1974) 1229 [Sov. J. Nucl. Phys. {\bf 20} (1975) 644]; 
S. Coleman, Phys. Rev. {\bf D15} (1977) 2929.

\bibitem{ak3} A. Kusenko, Phys. Lett. {\bf B377} (1996) 245 (hep-ph/9509275). 

\bibitem{cd}  J.~A.~Casas and S.~Dimopoulos, CERN-TH-96-116 
(hep-ph/9606237). 

\bibitem{bbc} H.~Baer, M.~Brhlik and D.~Castano, FSU-HEP-960801
(hep-ph/9607465).  

\bibitem{w} S.~Weinberg, Phys. Rev. Lett. {\bf 48} (1982) 1776.

\bibitem{cdl}  S.~Coleman and F.~De~Luccia, Phys. Rev. {\bf D21} 
(1980) 3305. 

\bibitem{r} I.~Dasgupta, B.~A.~Dobrescu and L.~Randall, BUHEP-96-25,
MIT-CTP-2555 (hep-ph/9607487). 

\bibitem{w_r} S.~Weinberg, Rev. Mod. Phys. {\bf 61} (1989) 1. 

\bibitem{a} L.~Abbott, Phys. Lett. {\bf B150} (1985) 427. 

\bibitem{wtn} E.~Witten, Mod. Phys. Lett. {\bf A10} (1995) 2153. 

\bibitem{lpi} S. Weinberg, Phys. Rev. {\bf D9} (1974) 3357;
R.~N.~Mohapatra and G.~Senjanovi\'{c}, Phys. Rev. Lett. {\bf 42} (1979)
1651; Phys. Rev. {\bf D20} (1979) 3390; P.~Langacker and S.-Y.~Pi, Phys.
Rev. Lett. {\bf 45} (1980) 1;  G.~Dvali and K.~Tamvakis, 
CERN-TH-96-45 (hep-ph/9602336);  A.~Melfo and G.~Senjanovi\'{c}, IC/96/76
(hep-ph/9605284). 

\bibitem{m} M.~Mangano,  Phys. Lett. {\bf B147} (1984) 307.

\bibitem{ce} D.~Comelli and J.~R.~Espinosa, DESY~96-114, FTUV/96-37 
(hep-ph/9606438). 

\bibitem{rrv} A.~Riotto, E.~Roulet and I.~Vilja, FERMILAB-PUB-96-170-A
(hep-ph/9607403). 

\bibitem{linde} L.~Kofman, A.~Linde and A.~A.~Starobinskii, Phys. Rev. Lett.
{\bf 73} (1994) 3195.

\bibitem{prefactor} C. G. Callan and S. Coleman, Phys. Rev. {\bf D16}
(1977) 1762;  A. Kusenko, Phys. Lett. {\bf B358} (1995) 47; A.~Kusenko, K.~Lee
and E.~J.~Weinberg, to appear.  

  
\bibitem{ak1} A. Kusenko, Phys. Lett. {\bf B358} (1995) 51. 

\bibitem{ccb} L.~Alvarez-Gaume, J.~Polchinski and M.~Wise, Nucl. Phys.
{\bf B221} (1983) 495; J.~M.~Frere, D.~R.~T.~Jones and S.~Raby, Nucl.
Phys. {\bf B222} (1983) 11; M.~Drees, M.~Gluck and K.~Grassie, 
Phys. Lett. {\bf B157} (1985) 164; J.~F.~Gunion, H.~E.~Haber and M.~Sher,
Nucl. Phys.  {\bf B306} (1988) 1; H.~Komatsu, Phys. Lett. {\bf B215} (1988)
323; P.~Langacker and N.~Polonsky, Phys. Rev. {\bf D 50} (1994) 2199;  
 A.~J.~Bordner, KUNS-1351 (hep-ph/9506409); J.~A.~Casas, A.~Lleyda and
C.~Mu\~noz, Nucl. Phys. {\bf B471} (1996)~3. 

\end{thebibliography}
\end{document}